\documentstyle[12pt]{article}
\begin{document}
\tolerance=5000
\def\be{\begin{equation}}
\def\ee{\end{equation}}
\def\bea{\begin{eqnarray}}
\def\eea{\end{eqnarray}}
\def\nn{\nonumber \\}
\def\cF{{\cal F}}
\def\det{{\rm det\,}}
\def\Tr{{\rm Tr\,}}
\def\tr{{\rm tr\,}}
\def\e{{\rm e}}
\def\etal{{\it et al.}}
\def\erp2{{\rm e}^{2\rho}}
\def\erm2{{\rm e}^{-2\rho}}
\def\er4{{\rm e}^{4\rho}}

\  \hfill 
\begin{minipage}{3.5cm}
NDA-FP-41 \\
November 1997 \\
\end{minipage}

\ 

\vfill

\begin{center}

{\large\bf Conformal anomaly for dilaton coupled 
electromagnetic field}

\vfill

{\sc Shin'ichi NOJIRI$^{\clubsuit}$}\footnote{
e-mail : nojiri@cc.nda.ac.jp} and 
{\sc Sergei D. ODINTSOV$^{\spadesuit}$}\footnote{
e-mail : odintsov@quantum.univalle.edu.co, \\ odintsov@galois.univalle.edu.co, 
odintsov@kakuri2-pc.phys.sci.hiroshima-u.ac.jp}

\vfill

{\sl $\clubsuit$ 
Department of Mathematics and Physics \\
National Defence Academy, 
Hashirimizu Yokosuka 239, JAPAN}

\ 

{\sl $\spadesuit$ 
Tomsk Pedagogical University, 634041 Tomsk, RUSSIA \\
and \\
Dep.de Fisica, Universidad del Valle, 
AA25360, Cali, COLOMBIA \\
}

\vfill

{\bf ABSTRACT}

\end{center}

The derivation of the conformal anomaly for 
dilaton coupled electromagnetic field in curved space 
is presented.  The models of this sort naturally appear 
 in stringy gravity or after spherical reduction of 
multidimensional Einstein-Maxwell theory. It is shown that unlike 
the case of minimal vector in curved space or dilaton coupled 
scalar the anomaly induced effective action cannot be derived. 
The reason is the same why anomaly induced effective action 
cannot be constructed for interacting theories (like QED) 
in curved space.

\newpage

\noindent{\bf 1.Introduction.}
In the study of string theory the lower energy 
 4D string effective action  maybe presented as following
\be
\label{I1}
S=\int d^4x \sqrt{-g}\left[ R + 4 (\nabla \phi)^2 + 
F_{\mu\nu}^2\right]\e^{-2\phi}
\ee
where $\phi$ is dilaton, $F_{\mu\nu}=\nabla_\mu A_\nu 
- \nabla_\nu A_\mu$, $A_\mu$ is electromagnetic field. 
 Investigating of string gravity (\ref{I1}) the quantum effects 
of electromagnetic field with the lagrangian:
\be
\label{I2}
L=-{1 \over 4}f(\phi)F_{\mu\nu}F^{\mu\nu}
\ee
maybe dominant in some regions,especially if we consider 
generalization of above model with N vectors and apply large N expansion. 
 From another point if we 
start from usual Einstein-Maxwell gravity in $D$-dimensions,
we can do spherical reduction to the space 
$R_4\times S_{D-4}$ where $R_4$ is an arbitrary $4D$ curved 
space.
Then the reduced action becomes again of the form (\ref{I1})
(maybe with change of some terms and some 
coefficients) where the radius of 
$S_{D-4}$ plays the role of dilatonic function.

Hence, the study of dilaton coupled electromagnetic field 
with the Lagrangian (\ref{I1}) which describes conformally 
invariant theory maybe of interest in different respects.
In the present letter we calculate the conformal anomaly 
for dilaton coupled vector (\ref{I2}) and discuss the 
problems which appear in the attempt to define the 
correspondent anomaly induced action.
Note that recently trace anomaly and anomaly induced 
action for dilaton coupled matter in two dimensions
has been discussed in Refs.\cite{E}, \cite{H}, \cite{NO} 
and \cite{Ku}.

\ 

\noindent{\bf 2.One-loop effective action and conformal anomaly 
for dilaton coupled electromagnetic field.}
The study of conformal anomaly \cite{DDI}, and its 
applications attracts a lot of attention (for a 
 recent review, see \cite{D}).  Among the most 
well-known applications one can list 
particle creation and Hawking radiation, 
the construction of non-singular Universe with back-reaction 
of quantum matter, anti-evaporation of black holes, etc. 
The conformal anomaly for 
electromagnetic field has been found in refs.\cite{BC,CD}. 
Our purpose in this section will be the calculation of 
conformal anomaly for dilaton coupled vector (or 
electromagnetic) field.

The initial Lagrangian of the theory has the following form;
\be
\label{I}
L=-{1 \over 4}f(\phi)F_{\mu\nu}F^{\mu\nu}
\ee
where  
$A_\mu$ is quantum vector field. We consider quantum theory 
with action (\ref{I}) in an external classical gravitational 
field. Note that electromagnetic field is non-minimally 
coupled with the external classical dilaton function 
$f(\phi)$ where $\phi$ is dilaton.

One can show that four dimensional theory with the 
Lagrangian (\ref{I}) is conformally invariant. 
Adding to the Lagrangian (\ref{I}) gauge-fixing 
Lagrangian $L_{gf}$:
\be
\label{II}
L_{gf}=-{1 \over 2}f(\phi)(\nabla_\nu A^\nu)^2\ ,
\ee
one can easily obtain
\be
\label{III}
L+L_{gf}={1 \over 2}f(\phi)A_\alpha \hat H^\alpha_\nu A^\nu\ .
\ee
Here 
\bea
\label{IV}
\hat H^\alpha_\nu &=& \delta^\alpha_\nu \Box - R^\alpha_\nu 
+ 2 [\hat h^\alpha_\nu ]^\mu \nabla_\mu\ , \nn
\left[ \hat h^\alpha_\nu \right]^\mu &=&{1 \over 2f}
\left\{\delta^\alpha_\nu (\nabla^\mu f) 
- g^{\alpha\mu}(\nabla_\nu f)
+ \delta^{\mu}_\nu(\nabla^\alpha f)\right\}\ .
\eea
Integrating over quantum vectors and taking into 
account the ghost contribution which is the same as 
in theory with $f=1$, one can get the one-loop effective 
action (its divergent part):
\be
\label{V}
\Gamma^{(1)}= \Gamma^{(1)}_{A_\mu} + \Gamma^{(1)}_{ghost}
=-{i \over 2}\Tr \ln \hat H^\alpha_\nu 
+i \Tr \ln \Box \ .
\ee
Here second term is ghost contribution.
In the dimensional regularization, one obtains
\be
\label{VI}
\Gamma^{(1)}={1 \over (n-4)}\int d^4 x \sqrt{-g}b_4 
\ee
where $b_4$ is $b_4$-coefficient of Schwinger-De Witt 
expansion.
Note that $\Tr \ln f(\phi)$ does not give the contribution 
to $\Gamma^{(1)}$ in frames of dimensional regularization.
For general algorithm of the calculation (\ref{VI}) one
can consult \cite{BOS}. 
According to this algorithm, 
$-{i \over 2}\Tr \ln \hat H^\alpha_\nu$ can be calculated as
\bea
\label{VII}
\Gamma^{(1)}_{A_\mu} &=& {1 \over (4\pi)^2(n-4)} 
\int d^4x \sqrt{-g}\tr \left\{ {1 \over 2}\hat P^2 
+ {1 \over 12}\hat S_{\alpha\beta}\hat S^{\alpha\beta} \right.
\nn
&& \left. {1 \over 6}\Box \hat P + {\hat 1 \over 180}
\left(R_{\mu\nu\alpha\beta}R^{\mu\nu\alpha\beta}
- R_{\mu\nu}R^{\mu\nu} + \Box R \right)\right\}
\eea
where trace is taken over vector space, i.e., 
$\tr\hat 1=4$.
For the operator $\hat H^\alpha_\nu$ we get
\bea
\label{VIII}
\hat P^\alpha_\nu &=& - R^\alpha_\nu 
+ \delta^\alpha_\nu {R \over 6}
- \nabla_\mu [\hat h^\alpha_\nu]^\mu
- [\hat h^\alpha_\beta]^\mu[\hat h^\beta_\nu]^\mu \nn
&=&- R^\alpha_\nu 
+ \delta^\alpha_\nu {R \over 6}
-{\delta^\alpha_\nu \over 2f}\Box f 
+ {1 \over 2f^2}[\delta^\alpha_\nu (\nabla_\mu f)
(\nabla^\mu f) + (\nabla^\alpha f)
(\nabla_\nu f)]\ , \nn
\hat S_{\alpha\beta}&\equiv& [\hat S_{\mu\nu}]_{\alpha\beta} 
\nn
&=&(\nabla_\beta\nabla_\alpha 
-\nabla_\alpha \nabla_\beta) \hat 1 
+ \nabla_\beta [\hat h_{\mu\nu}]_\alpha \nn
&&- \nabla_\alpha [\hat h_{\mu\nu}]_\beta 
+ [\hat h_{\mu\tau}]_\beta[\hat h^\tau_\nu]_\alpha
- [\hat h_{\mu\tau}]_\alpha[\hat h^\tau_\nu]_\beta 
\nn
&=& R_{\mu\nu\beta\alpha}-{1 \over 4f^2}
\{-g_{\alpha\mu}(\nabla_\beta f)(\nabla_\nu f)
+g_{\alpha\nu}(\nabla_\beta f)(\nabla_\mu f) \nn
&& +g_{\beta\mu}(\nabla_\alpha f)(\nabla_\nu f)
-g_{\beta\nu}(\nabla_\alpha f)(\nabla_\mu f) \nn
&& +g_{\mu\beta}g_{\nu\alpha}(\nabla_\tau f)(\nabla^\tau f)
-g_{\mu\alpha}g_{\nu\beta}(\nabla_\tau f)(\nabla^\tau f) \nn
&& +{1 \over 2f}[g_{\alpha\nu}(\nabla_\beta\nabla_\mu f)
-g_{\alpha\mu}(\nabla_\beta\nabla_\nu f) \nn
&& +g_{\beta\mu}(\nabla_\alpha\nabla_\nu f)
-g_{\beta\nu}(\nabla_\alpha\nabla_\mu f)]\ .
\eea
Having explicit expressions (\ref{VIII}) for the 
operators $\hat S_{\alpha\beta}$, $\hat P$, one can find;
\bea
\label{IX}
{1 \over 2}\tr\hat P^2&=&{1 \over 2}R^{\alpha\beta}
R_{\alpha\beta}-{1 \over 9}R^2
+{R \over 6f}\Box f + {1 \over 2f^2}(\Box f)^2 \nn
&&-{R^{\nu\alpha} \over 2f^2}(\nabla_\nu f)(\nabla_\alpha f) 
-{R \over 12 f^2}(\nabla_\mu f)(\nabla^\mu f) \nn
&&-{5 \over 4f^3}(\Box f)(\nabla_\mu f)(\nabla^\mu f) \nn
&& +{7 \over 8f^4}(\nabla_\mu f)(\nabla^\mu f) 
(\nabla_\alpha f)(\nabla^\alpha f) \ ,\nn
{1 \over 6}\tr\Box\hat P &=& -{1 \over 18}\Box R 
- {1 \over 3}\Box\left({1 \over f}\Box f\right) 
+ {5 \over 12}\Box \left[{1 \over f^2}
(\nabla_\alpha f)(\nabla^\alpha f) \right] \ ,\nn
{1 \over 12}\hat S_{\alpha\beta}\hat S^{\alpha\beta}
&=&-{1 \over 12}R_{\mu\nu\alpha\beta}R^{\mu\nu\alpha\beta}
+{1 \over 6f^2}R_{\mu\beta}(\nabla^\mu f)(\nabla^\beta f) 
+{1 \over 12f^2}R(\nabla_\mu f)(\nabla^\mu f) \nn
&& -{1 \over 3f}R_{\alpha\beta}(\nabla^\beta\nabla^\alpha f) 
-{5 \over 16f^2} (\nabla_\beta f)(\nabla^\beta f) 
(\nabla_\mu f)(\nabla^\mu f) \nn
&& +{1 \over 6f^3}(\nabla^\mu f)(\nabla^\nu f) 
(\nabla_\mu\nabla_\nu f) 
+ {1 \over 3f^3}(\Box f)(\nabla_\beta f)(\nabla^\beta f) \nn
&& -{1 \over 6f^2}(\nabla_\alpha\nabla_\beta f)
(\nabla^\alpha\nabla^\beta f)
-{1 \over 12f^2}(\Box f)(\Box f)
\eea
On the same time, for the second (ghost) term in Eq.(\ref{V}), 
we get:
\be
\label{X}
\hat S_{\alpha\beta}=0\ ,\ \ \ \hat P={R \over 6}\ .
\ee
Hence, using Eq.(\ref{VII}) we will find 
\bea
\label{XI}
\Gamma^{(1)}_{ghost}&
=&{1 \over (4\pi)^2 (n-4)}
\int d^4x \sqrt{-g}\left\{-{1 \over 90}R_{\mu\nu\alpha\beta}^2
+{1 \over 90}R_{\mu\nu}^2 \right. \nn
&& \left. -{R^2 \over 36} -{1 \over 15}\Box R\right\} \nn
&=&{1 \over (n-4)}\int d^4 x \sqrt{-g}b_4^{ghost} \ .
\eea
As one can see this is standard expression for the 
ghost contribution to one-loop effective action, it 
does not depend on dilaton.

The one-loop effective action due to vectors (\ref{VII})  may be found as

\bea
\label{XIII}
\Gamma^{(1)}_{A_\mu}&=&{1 \over (4\pi)^2 (n-4)}
\int d^4x \sqrt{-g}\left\{
-{11 \over 180}R_{\mu\nu\alpha\beta}^2
+{43 \over 90}R_{\mu\nu}^2 -{1 \over 9}R^2 \right. \nn
&& - {1 \over 30}\Box R -{1 \over 3}\Box\left({1 \over f}
\Box f\right) + {5 \over 12}\Box \left[{1 \over f^2} 
(\nabla_\alpha f)(\nabla^\alpha f) 
\right]
\nn
&& -{1 \over 3f^2}
R_{\mu\nu}(\nabla^\mu f)(\nabla^\nu f) 
-{1 \over 3f}R_{\mu\nu}(\nabla^\mu\nabla^\nu f) 
+{R \over 6f}(\Box f) \nn
&& +{9 \over 16f^4} (\nabla_\mu f)(\nabla^\mu f) 
(\nabla_\alpha f)(\nabla^\alpha f) 
+{1 \over 6f^3}(\nabla^\beta f)(\nabla^\nu f) 
(\nabla_\beta\nabla_\nu f) \nn
&& - {11 \over 12f^3}(\nabla^\mu f)(\nabla_\mu f)(\Box f)
 -{1 \over 6f^2}(\nabla_\alpha\nabla_\beta f)
(\nabla^\alpha\nabla^\beta f) \nn
&& \left. +{5 \over 12f^2}(\Box f)(\Box f)\right\} \nn
&=&{1 \over (n-4)}\int d^4 x \sqrt{-g}b_4^{vector}  \ .
\eea
Hence, we found the one-loop effective action due to 
dilaton coupled vectors. 
The total one-loop effective action $\Gamma^{(1)}$ is 
given by sum of (\ref{XI}) and (\ref{XIII}).

 From the expression for $\Gamma^{(1)}$ one can easily 
get conformal anomaly:
\be
\label{XIV}
T=b_4 =b_4^{ghost}+b_4^{vector}\ .
\ee
The first four terms give the well-known conformal anomaly 
of electromagnetic  field \cite{BC,CD}.
We have to note that using another regularization (like 
zeta-regularization) may slightly alter the coefficients 
of total derivative terms in (\ref{XIV}), like already 
happened in this case in the absence of dilaton 
\cite{BC,CD}.

For completeness, we write below the trace anomaly for dilaton 
coupled conformal scalar with the Lagrangian;
\be
\label{XV}
L=\varphi f(\phi)\left(\Box - {1 \over 6}R\right)\varphi\ .
\ee
In this case, we get \cite{NO}
\bea
\label{XVI}
T&=&{1 \over (4\pi)^2}
\left\{ {1 \over 32}{[(\nabla f) (\nabla f)]^2 \over f^4}
+{1 \over 24}\Box \left(
{(\nabla f) (\nabla f) \over f^2}\right) \right. \nn
&& \left. + {1 \over 180}
\left( R_{\mu\nu\alpha\beta}^2 - R_{\mu\nu}^2 
+ \Box R \right)\right\}
\eea
If one considers the system consisting of 
$n$ dilaton coupled conformal scalars and $m$ dilaton 
coupled vectors then the total conformal anomaly of the 
system is given by:
\be
\label{XVII}
T=nT(17)+mT(15)
\ee
 Hence, we found the explicit expression for dilaton coupled vector 
conformal anomaly.
It would be of interest to study the structure of 
this anomaly extending results of ref.\cite{DS}.

\ 

\noindent{\bf 3.Anomaly induced effective action.}
With the help of conformal anomaly one can construct the anomaly 
induced effective action on the same way as in 
ref.\cite{R}.
First of all, we will rewrite $f$-independent terms 
of conformal anomaly (\ref{XIV}) in a slightly 
different form:
\be
\label{XVIII}
T={1 \over (4\pi)^2}
\Bigl\{b\left(F+ {2 \over 3}\Box R \right) 
+ b'G + b'' \Box R +  \cdots
\ee
where $b={1 \over 10}$, $b'=-{31 \over 180}$ and in 
our model $b''=-{1 \over 6}$.
Note, however, that in principle $b''$ corresponds to 
arbitrary parameter because $\Box R$ is the variation of 
local action:
$\sqrt{-g}\Box R=-{1 \over 6}g^{\mu\nu}{\delta \over 
\delta g^{\mu\nu}}\int d^4x\sqrt{-g}R^2$.
Hence, $b''$ may be always changed by the addition of 
finite counterterms to gravitational effective action.

Let the metric has the form
\be
\label{XIX}
g_{\mu\nu}=\e^{2\sigma}\bar g_{\mu\nu}\ .
\ee
In this case for Ricci tensor and Ricci scalar we get
\bea
\label{XX}
R&=&\e^{-2\sigma}[\bar R - 6\bar{\bar\Box}\sigma 
-6(\bar\nabla_\mu\sigma)(\bar\nabla^\mu\sigma)]\ ,\nn
R_{\mu\nu}&=&[\bar R_{\mu\nu}
-2\bar\nabla_\mu\bar\nabla_\nu\sigma 
-\bar g_{\mu\nu}\bar{\bar\Box}\sigma  \nn
&& +2(\bar\nabla_\nu\sigma)(\bar\nabla_\mu\sigma)
-2\bar g_{\mu\nu}\bar\nabla^\tau\sigma\bar\nabla_\tau\sigma]\ .
\eea
One can write now the conformal anomaly (\ref{XIV}), 
(\ref{XVIII}) in the following form;
\be
\label{XXI}
T={1 \over \sqrt{-g}}{\delta \over \delta \sigma}
W[\sigma]
\ee
where $W[\sigma]$ is some unknown anomaly induced effective 
action which should be found after integration of 
Eq.(\ref{XXI}).
 Substituting conformally transformed curvature 
tensors and metric to conformal anomaly, we find
\bea
\label{XXII}
\sqrt{-g}T&=&{\sqrt{-\bar g} \over (4\pi)^2}
\left\{b\bar F + b'\left(\bar G -{2 \over 3}\bar{\bar\Box}\bar R
\right) + 4b'\bar\Delta\sigma + \left[ b''+{2 \over 3}(b+b')\right]
\Box R \e^{4\sigma} \right. \nn
&& -{1 \over 3}\left[ -{2(\bar{\bar\Box}\sigma) \over f}
(2(\bar\nabla_\mu\sigma)(\bar\nabla^\mu f)+\bar{\bar\Box} f) \right.
\nn
&& -2(\bar\nabla_\mu\sigma)\bar\nabla^\mu\left({1 \over f}
(2(\bar\nabla_\mu\sigma)(\bar\nabla^\mu f) + \bar{\bar\Box} f)\right) 
\left. +\bar{\bar\Box}\left({1 \over f}(2(\bar\nabla_\mu\sigma)
(\bar\nabla^\mu f) + \bar{\bar\Box} f)\right)\right] \nn
&& +{5 \over 12}\left[\bar{\bar\Box} \left((\bar\nabla_\mu f)
(\bar\nabla^\mu f)\right) 
-2{(\bar\nabla_\mu f)(\bar\nabla^\mu f) \over f^2}
(\bar{\bar\Box}\sigma) \right. \nn
&& \left.-2 (\bar\nabla^\mu\sigma)\bar\nabla_\mu
\left[{(\bar\nabla_\mu f)(\bar\nabla^\mu f) \over f^2}
\right]\right] \nn
&& + \left[\bar R_{\mu\nu}-2\bar\nabla_\mu\bar\nabla_\nu\sigma 
-\bar g_{\mu\nu}\bar{\bar\Box}\sigma + 
2(\bar\nabla_\nu\sigma)(\bar\nabla_\mu\sigma)
-2\bar g_{\mu\nu}(\bar\nabla^\tau\sigma)(\bar\nabla_\tau\sigma) 
\right] \nn
&& \times \left[-{1 \over 3f^2}(\bar\nabla^\mu f)
(\bar\nabla^\nu f)-{1 \over 3f}\{(\bar\nabla^\mu\bar\nabla^\nu
 f) - \bar g^{\mu\tau}(\bar\nabla^\nu\sigma)(\bar\nabla_\tau 
f) \right. \nn
&& \left.-\bar g^{\nu\tau}(\bar\nabla^\mu\sigma)
(\bar\nabla_\tau f) 
+\bar g^{\mu\nu}(\bar\nabla^\tau\sigma)
(\bar\nabla_\tau f) \}\right] \nn
&& + {1 \over 6f}[2(\bar\nabla_\mu\sigma)(\bar\nabla^\mu f)
+ \bar{\bar\Box} f]\times[\bar R - 6\bar{\bar\Box}\sigma 
- 6(\bar\nabla_\mu\sigma)(\bar\nabla^\mu\sigma)] \nn
&& + {9 \over 16f^4}[(\bar\nabla_\mu f)^2]^2 
+ {1 \over 6f^3}(\bar\nabla^\beta f)(\bar\nabla^\nu f)
\left(\bar\nabla_\beta\bar\nabla_\nu f 
-(\bar\nabla_\beta\sigma)(\bar\nabla_\nu f) \right. \nn
&& \left. -(\bar\nabla_\beta f)(\bar\nabla_\nu\sigma)
+\bar g_{\beta\nu}(\bar\nabla^\tau\sigma)(\bar\nabla_\tau f)\right) 
-{11 \over 12 f^3}(\bar\nabla_\mu f)^2
[2(\bar\nabla_\nu\sigma)(\bar\nabla^\nu f) 
+ \bar{\bar\Box} f ] \nn
&& -{1 \over 6f^2}\left[\bar\nabla_\alpha\bar\nabla_\beta f 
- (\bar\nabla_\alpha f)(\bar\nabla_\beta\sigma) 
- (\bar\nabla_\alpha\sigma)(\bar\nabla_\beta f)
+ \bar g_{\alpha\beta}(\bar\nabla^\tau\sigma)
(\bar\nabla_\tau f)\right]  \nn
&& \times[\bar\nabla^\alpha\bar\nabla^\beta f 
- (\bar\nabla^\alpha f)(\bar\nabla^\beta\sigma)
- (\bar\nabla^\alpha\sigma)(\bar\nabla^\beta f)
+ \bar g^{\alpha\beta}(\bar\nabla^\tau\sigma)
(\bar\nabla_\tau f)] \nn
&& \left.+{5 \over 12f^2}[2(\bar\nabla_\mu\sigma)
(\bar\nabla^\mu f) + \bar{\bar\Box} f]^2\right\}\ .
\eea
Here 
$\bar\Delta=\bar{\bar\Box}^2 + 2\bar R^{\mu\nu}
\bar\nabla_\mu\bar\nabla_\nu 
-{2 \over 3}\bar R \bar{\bar\Box} 
+{1 \over 3}(\bar\nabla^\mu\bar R)\bar\nabla_\mu$.
It is also easier to keep fourth term in (\ref{XXII}) 
to be non-transformed.

Now, using Eq.(21), one can integrate expression (\ref{XXI}) 
with conformal anomaly (\ref{XXII}) in order to get the 
explicit effective action $W$
\bea
\label{XXIII}
W&=&b\int d^4x \sqrt{-\bar g}\bar F \sigma 
+ b'\int d^4x \sqrt{-\bar g}
\left\{2\sigma\bar\Delta\sigma + \left(\bar G 
- {2 \over 3}\bar{\bar\Box} R\right)\sigma\right\} \nn
&&-{1 \over 12}\left[b''+{2 \over 3}(b+b')\right]
\int d^4x\sqrt{-\bar g}\left[ \bar R - 6\bar{\bar\Box}\sigma 
-6(\bar\nabla_\mu\sigma)(\bar\nabla^\mu\sigma)\right]^2 \nn
&&+\int d^4x\sqrt{-\bar g}\left[ 
-{1 \over 3}\left\{ \sigma\bar{\bar\Box}^2 f 
+{1 \over f}(\bar\nabla_\mu\sigma)
(\bar\nabla^\mu\sigma)\bar{\bar\Box} f \right.\right. \nn
&& \left. + {4 \over 3f}(\bar\nabla_\mu\sigma)(\bar\nabla^\mu\sigma)
(\bar\nabla_\mu\sigma)(\bar\nabla^\mu f) 
\right\} \nn
&& +{5 \over 12}\left\{
\sigma\bar{\bar\Box} \left((\bar\nabla_\mu f)
(\bar\nabla^\mu f)\right) 
+{1 \over f^2}(\bar\nabla^\mu\sigma)
(\bar\nabla_\mu\sigma)
(\bar\nabla_\mu f)(\bar\nabla^\mu f) \right\} \nn
&& +\bar R_{\mu\nu}\left\{-{1 \over 3f^2}(\bar\nabla^\mu f)
(\bar\nabla^\nu f)-{1 \over 3f}
(\bar\nabla^\mu\bar\nabla^\nu f)\right\}\sigma \\
&& +{1 \over 6f}\sigma\bar{\bar\Box f}\bar R
+ {9 \over 16f^4}\sigma[(\bar\nabla_\mu f)^2]^2 
+ {1 \over 6f^3}\sigma(\bar\nabla^\beta f)(\bar\nabla^\nu f)
\bar\nabla_\beta\bar\nabla_\nu f \nn
&& \left . -{11 \over 12 f^3}\sigma(\bar\nabla_\mu f)^2
\bar{\bar\Box} f 
-{1 \over 6f^2}\sigma\bar\nabla_\alpha\bar\nabla_\beta f 
\bar\nabla^\alpha\bar\nabla^\beta f 
+{5 \over 12f^2}\sigma(\bar{\bar\Box} f)^2 \right] + \cdots \nonumber
\eea
However, several terms in the trace anomaly (\ref{XXII}) cannot 
be integrated to give the effective action. It is not difficult 
to write explicitly all such terms if it is necessary.
 
 The reason why we cannot integrate some terms in conformal anomaly 
(\ref{XIV}) is the same as it was pointed out in the paper by 
Reigert \cite{R} for $R^2$-term.
I.e., the functional derivative of $\sqrt{-g}T(x)$ with 
respect to $\sigma(y)$ must be symmetric in $y$ and $x$ 
if anomaly induced action would exist. 
However it is not difficult to show that functional derivative of 
(\ref{XIV}) is not symmetric one for most of terms. 
Hence we conclude that anomaly induced action does not exist for 
dilaton coupled vector, i.e., already in the free theory.

On the contrary, anomaly induced action for free fields non-interacting 
with dilaton may be easily constructed. 
Only afer taking account of quantum field interaction $R^2$ 
term with the coefficient proportional to coupling constant
appeared. 
Only after appearence of this term anomaly induced action could not 
be constructed.
 However we see for dilaton coupled 4$D$ scalar anomaly induced action 
does exist\cite{NO}:
\bea
\label{XXV}
W&=&b\int d^4x \sqrt{-\bar g} \bar F\sigma 
+b'\int d^4x \sqrt{-g} \left\{2\sigma\bar\Delta \sigma
 + \left(\bar G-{2 \over 3}\bar{\bar\Box} R\right)\sigma \right\} \nn
&& -{1 \over 12}\left(b'' + {2 \over 3}(b + b')\right)
\int d^4x \sqrt{-\bar g}\left[\bar R - 6 \bar{\bar\Box} \sigma 
- 6(\bar\nabla \sigma)(\bar\nabla \sigma) \right]^2 \nn
&& + \int d^4x \sqrt{-\bar g} \left\{ 
a_1 {[(\bar\nabla f) (\bar\nabla f)]^2 \over f^4}\sigma 
+a_2\bar{\bar\Box} \left({(\bar\nabla f) (\bar\nabla f) \over f^2} 
\right)\sigma \right. \nn
&& \hskip 2cm \left. 
+a_2{(\bar\nabla f) (\bar\nabla f) \over f^2} [(\bar\nabla \sigma) 
(\bar\nabla \sigma )]\right\}
\eea
where
\be
\label{v8}
b={ 1 \over 120 (4\pi)^2 }\ ,\ \ 
b'=-{ 1 \over 360 (4\pi)^2 }\ ,\ \ 
a_1={ 1 \over 32 (4\pi)^2 }\ ,\ \ 
a_2={ 1 \over 24 (4\pi)^2 }\ .
\ee
One can investigate different cosmological applications of anomaly induced 
effective action (\ref{XXV}), like study of early Universe with 
back-reaction of dilaton coupled scalar matter, constructing of 
new versions of IR sector of dilatonic gravity, etc. 

\ 

\noindent{\bf 4.Conclusion}
We calculated conformal anomaly for dilaton coupled 
electromagnetic field. It is shown that it is impossible to 
construct anomaly induced effective action in this case.
Currently we do not understand the physical reason
  why it is impossible to integrate 
conformal anomaly. It could be that only in the case of dilaton 
coupled conformal supermultiplet the dangerous terms in (super) 
conformal anomaly cancel and anomaly induced effective action can be 
constructed. This question deserves futher study.

\ 

\noindent{\bf Acknoweledgments.} 
We would like to thank S. Ichinose for 
independent check of the derivation of conformal anomaly 
and R. Bousso and S. Hawking for helpful e-mail discussion.
The work of SDO has been partially supported by 
Universidad del Valle and
COLCIENCIAS(Colombia),Graduate College of Leipzig University (Germany)
and RFBR project No 96-02-16017(Russia).

\end{document}